\documentclass[twocolumn,showpacs,amsmath,amssymb]{revtex4}

\usepackage{graphicx}

\begin{document}

\title{Unusual polarization patterns in flat epitaxial ferroelectric nanoparticles}
\author{Ivan I. Naumov and Alexander M. Bratkovsky}
\affiliation{Hewlett-Packard Laboratories, 1501 Page Mill Road, Palo Alto, California
94304}
\date{\today }

\maketitle

\textbf{Interest in epitaxial ferroelectric nanoislands was growing rapidly
in recent years driven by their potential for devices, especially ultradense
memories \cite{scott, lichtensteiger,gruverman,alexe}. Recent advances in
the \textquotedblleft bottom-up" (self-assembly) nanometer scale techniques have opened up
the opportunities of fabricating high-quality epitaxial ferroelectric
nanoislands with extremely small thickness and lateral size on the order of
1 nm and 20 nm, respectively \cite{alexe,roelofs, seol,okaniwa,shimizu}. On
the other hand, recent emergence of powerful probes, such as piezoresponse
force microscopy (PFM), has enabled imaging of a local domain structure with
sub-10 nm resolution \cite{gruverman,alexe2,rudiger}. In spite of those
developments, a clear understanding of the polarization patterns in
epitaxial ferroelectric nanoislands is lacking, and some important
characteristics, like a critical lateral size for ferroelectricity, are not
yet established. Here, we perform \textit{ab-initio} studies of
non-electroded epitaxial Pb(Zr$_{0.5}$Ti$_{0.5}$)O$_{3}$ (PZT) and BaTiO$%
_{3} $ (BTO) nanoislands and show the existence of novel polarization
patterns driven by the misfit strains and/or anisotropy energy. The results
allow interpretation of the data and design of the ferroelectric
nanostructures with tailored response to external field}.

Geometrically, nanoislands or flat nanoparticles represent a class of
systems that bridge the gap between 0D nanodots and 2D ultra-thin films.
Compared with thin films, they have free-standing side walls that tend to
suppress the formation of uniform in-plane polarization because of appearing
depolarizing field, like in ferroelectric \cite{kanzig1, kanzig2}  or ferromagnetic
particles \cite{hubert}. On the other hand, relative to the (confined in all
three dimensions) nanodots, they have large aspect ratio and likely to
behave similarly to thin films when the polarization is out-of-plane. These
observations lead one to expect that ferroelectric nanoislands should
exhibit some kind of a \textquotedblleft particle-to-thin film" crossover
behavior and related novel effects depending on the aspect ratio and the
type of bulk polarization ordering (which is different in PZT\ compared to
BTO\ crystals).

Free-standing ferroelectric nanodots usually lose stability with respect to
a vortex ground state, in which the polarization curls around some vortex
core(s). Such an ordering can be characterized, in simplest single vortex
cases, by a \emph{toroidal} moment or a moment of polarization {\cite{naumov,
ponom1, ponom2,prosand}. This is analogous, to some extent, to the magnetic
vortices in magnetic particles \cite{hubert}, but there are also qualitative
differences due to electrostatic interactions and inhomogeneity energy in
ferroelectric particles being larger than the magnetic analogues by many
orders of magnitude. Note also that the symmetry of the bulk cubic crystals
of BTO\ or PZT\ does not allow for a \textquotedblleft bulk"
 toroidal moment \cite{gorbatsevich},
such an ordering in the present case is entirely due to presence of boundaries of the
particles
and the accompanying it depolarizing/demagnetizing field. The moment of polarization appears
continuously at the critical point and may be regarded as an order parameter
for such a transition in an island \cite{naumov}.}
\begin{figure*}[tbp]
\centering
\includegraphics[height=95mm]{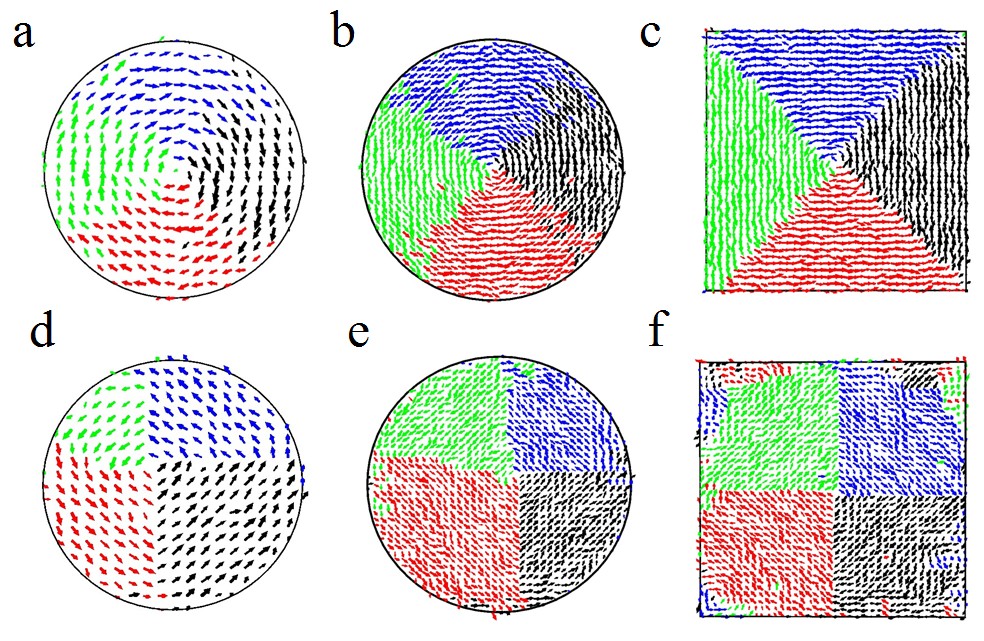}
\caption{Low-temperature polarization distributions on the central $z$-plane
in PZT (a,b,c) and BTO (d,e,f) nanoparticles in free-standing states. (a,d)--
19$\times$6 nanodisks, (b,e)-- 39$\times$10 nanodisks and (c,f)--39$\times$39%
$\times$10 particles with square footprints. Shown in red, blue, green and
black are the dipoles directed predominantly along [$\pm$100] and [0$\pm$10]
in PZT and along [$\pm$1$\pm$10] in BTO systems.}
\label{fig:fig1}
\end{figure*}

{We shall consider below the nanoislands of cubic perovskite materials. The
epitaxial thin films of such materials become uniaxial under compressive
strain with an easy axis perpendicular to the film, and usually form a 180$%
^{0}$ }stripe (c-)domain{\ structure \cite%
{bl00,tinte,fong1,streiffer,fong2,kornev,wu1,wu2, bo1, bo2}. One can,
therefore, anticipate the flat ferroelectric nanoislands with large aspect ratio
to undergo an unusual vortex to stripe domains transformation with an
increasing compressive misfit strain. We show below that perovskite
nanoislands do indeed undergo such a transition, but only via one or two
\emph{intermediate}\textit{\ }}phases{\ where in-plane curling polarization
coexist with the 180$^{0}$ c-domains. Moreover, the domain phase does not
necessarily render the well known 180$^{0}$ stripe domain structure, but
also can take the form of a 180$^{0}$ \textit{tweed texture}, depending on
the anisotropy energy. We also demonstrate that stripe and tweed textures
evolve differently in applied external electric field.}

The simulated nanoparticles have circular and/or rectangular shapes and the
following dimensions: 19$\times 6$, 39$\times $10 for the former and 39$\times
$39$\times$10 for the latter (all dimensions are given in lattice parameters of
the cubic bulk phase $a$, which is 4.00\AA\ and 3.95\AA\ for PZT and BTO, respectively). The $z-$%
axis is selected along the (shortest)\ pseudocubic [001] direction normal to
the substrate. To simulate finite-temperature behavior of the particles, we
use the effective Hamiltonian
\begin{equation}
H\{\boldsymbol{u}_{i}\}=H^{b}\{\boldsymbol{u}_{i}\}+H^{s}\{\boldsymbol{u}%
_{i}\}-\sum_{i}Z^{\ast }\boldsymbol{u}_{i}\cdot \boldsymbol{E}_{0},
\label{eq:1}
\end{equation}%
whose primary variable is $\boldsymbol{u}_{i}$, the local soft mode in the
unit cell $i$ \cite{Heff, bellaiche}. The first part of the Hamiltonian, $%
H^{b}$ is the same as in infinite bulk materials: its
first-principles-derived parameters for BTO are given in \cite{Heff}, and
for PZT- in \cite{bellaiche}. The second part, $H^{s}$, represents the
corrections to the $H^{b}$ associated with the presence of free surface via
\textquotedblleft vacuum-local mode" and \textquotedblleft vacuum-inhomogeneous strain"
interactions \cite{fu,emad}. And, finally, the third term describes the
interaction of the local site dipoles $\boldsymbol{p}_{i}=Z^{\ast }%
\boldsymbol{u}_{i}$ with an external field $\boldsymbol{E}_{0}$ \cite{garcia}
($Z^{\ast }$ is the effective Born charge of the local mode $\boldsymbol{u}%
_{i}$). We should stress that the first-principles-derived Hamiltonian is
able to reproduce some nontrivial properties beyond the properties it was
fit to: phase diagram, occurrence of an exotic monoclinic phase in a small
range of Ti concentrations in bulk PZT, stripe c-domains in thin films with
compressive in-plane strains, etc. Typically, 40,000 Monte Carlo sweeps are
used to find the equilibrium dipole configuration at a fixed temperature.
Open-circuit boundary conditions are assumed, so that no screening charges
are taken into consideration. The effect of a substrate is imposed by fixing
the homogeneous in-plane strain in the whole island: $\epsilon
_{xx}=\epsilon _{yy}=\epsilon $, $\epsilon _{xy}=0$.

Our calculations predict that at low temperatures and in free-standing
states all the investigated nanoparticles transform into a vortex ground
state characterized by a nonzero toroidal moment $\boldsymbol{G}%
=(2N)^{-1}\sum_{i}\boldsymbol{r}_{i}\times \boldsymbol{p}_{i}$, where $N$ is
the total number of unit cells. In disk-shaped PZT particles (Fig.~1a,b),
the local dipoles rotate from cell to cell forming the classical vortex
patterns. But in a rectangular PZT 39$\times $39$\times $10 particle the
polarization aligns with the square boundary as much as possible, hence the
discontinuity lines (domain walls)\ form, and the resulting pattern can be
viewed as consisting of four domains separated by 90$^{0}$ domain walls
(Fig.~1c). This is in a clear analogy with ordering in magnetic particles
\cite{hubert}, anticipated for ferroelectric particles back in 1950s \cite
{kanzig1, kanzig2}. The polarization in each domain point along one of the four
pseudocubic [$\pm $100] and [0$\pm $10] directions, so that the domain walls
are parallel to the (1$\pm $10) crystallographic planes and pass through the
center of the particle. In BTO particles, regardless of their shape and
size, the dipoles tend to be directed along the [$\pm $1$\pm $10] directions
(Fig.~1d,e,f). As a result, two (100)-type 90$^{0}$ domain walls are
likely to be formed crossing each other at the geometric center of the
particles. The walls are rather fuzzy in bigger particles, especially in a
rectangular 39$\times $39$\times $10 dot where the dipoles are frustrated
near the lateral surfaces.

The different behavior of PZT and BTO free-standing nanoparticles can be
easily understood by looking at the anisotropy energy $\sum_{i}\gamma
_{i}(u_{ix}^{2}u_{iy}^{2}+u_{ix}^{2}u_{iz}^{2}+u_{iy}^{2}u_{iz}^{2})$ in the
effective Hamiltonian $H$ (its part $H^{b}$). Suppose that the amplitude $|u|
$ is fixed, then the anisotropy energy will approach its minimum either for $%
\mathbf{u}$ along the [100] direction ($\gamma >0$) or along the [111]
direction ($\gamma <0$). For PZT, the parameter $\gamma $ (associated with
 the dominating  Ti-centered unit cells) is positive \cite{bellaiche} -- this is
consistent with the fact that PZT in bulk form adopts a tetragonal structure
with an \textquotedblleft easy" direction of polarization along [100]. At the same time, in
BTO $\gamma<0$ \cite{Heff}, and this system in bulk has a rhombohedral
equilibrium structure with easy direction [111] and the equivalent bulk
diagonals. Let us compare now, for example, two 39$\times $39$\times $10 PZT
and BTO particles. In both cases, the effects of depolarizing field force
the dipoles to lie in the $x,y-$plane. In the case of the PZT particle,
however, the materials anisotropy factor is in compliance with the shape:
majority of the dipoles are oriented along the easy directions [100] and
[010] and at the same time they are parallel to the surface. The situation
is quite different for the BTO particles, where \emph{none} of the dipoles
can be directed along the easy [111] body diagonal: in this case the
majority of dipoles in the center region prefer to be oriented along [110]
instead of the [111] direction. Near the side surfaces, however, the
depolarizing field forces the dipoles to align with the square geometry,
which increases the anisotropy energy term. The competition between these
forces leads to the dipole frustration near the vertical walls and to the
formation of four additional vortices near the corners (Fig.~1f) in agreement
with the previous calculations for a 24$\times $24$\times $24 BTO nanodot
\cite{fu}.

\begin{figure}[tbp]
\centering
\includegraphics[scale=1.1]{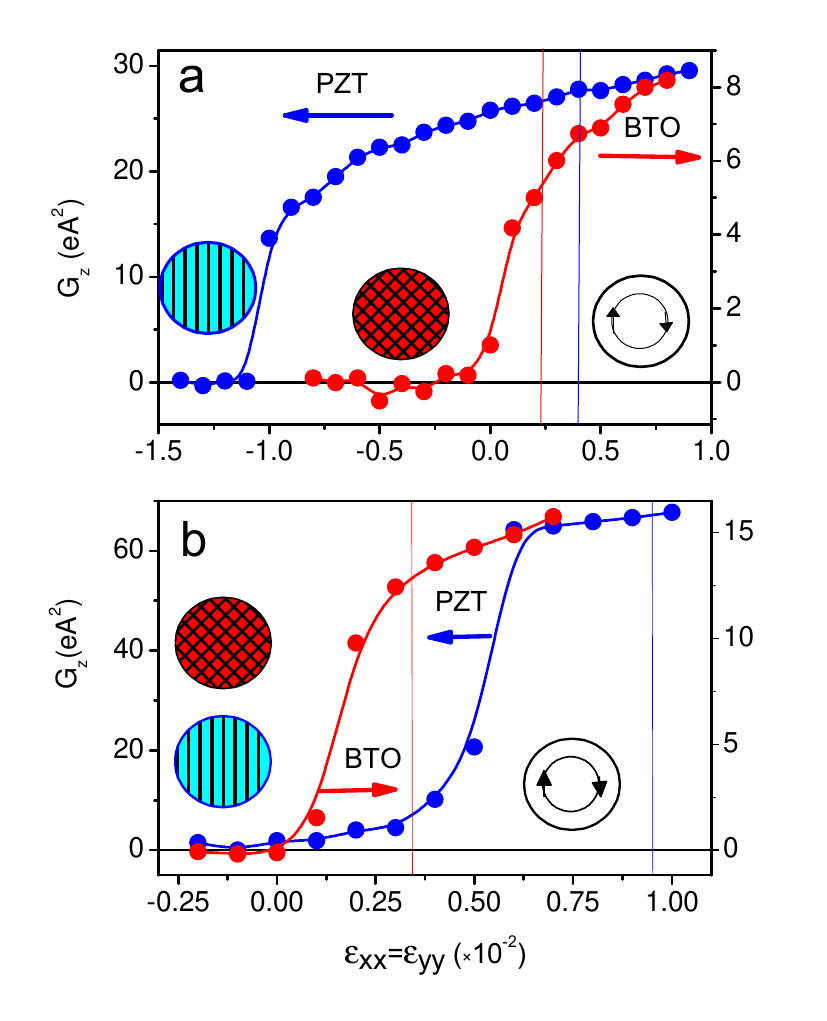}
\caption{The $z$-component of the toroidal moment as a function of strain in
19$\times$6 (a) and 39$\times$10 (b) nanoparticles. The strains are measured
relative to the theoretical LDA-calculated lattice constants in bulk cubic
structures. Note that in free standing states the particles are
characterized by non-zero \textquotedblleft residual" strains marked by vertical lines. The
residual strain increases in passing from 19$\times$6 to 39$\times$10
systems, especially in PZT.}
\label{fig:fig2}
\end{figure}
The tensile strains stabilize the vortex states further, as seen from Fig.
2, increasing the $z$-component of the toroidal moment. On the contrary,
compressive strains enhance the out-of-plane polarization, decrease $G_{z}$
and eventually lead, as analysis below shows, to a 180$^{0}$ multi-domain
structure running through the entire thickness of the islands. It is
remarkable that in the case of BTO particles the critical strain associated
with this transition is very small, approximately zero, $\approx $0 $\%$. In
the PZT case, the critical strain dramatically decreases from 1.2 $\%$ to $%
\approx $0 $\%$ when one increases the diameter from 19 to 39. The bigger
PZT particle in a vortex state can tolerate the compressive strains less
because its optimal in-plane averaged lattice parameter is noticeably larger
than that corresponding to the bulk cubic phase (Fig. 2b).

\begin{figure}[tbp]
\centering
\includegraphics[height=73mm]{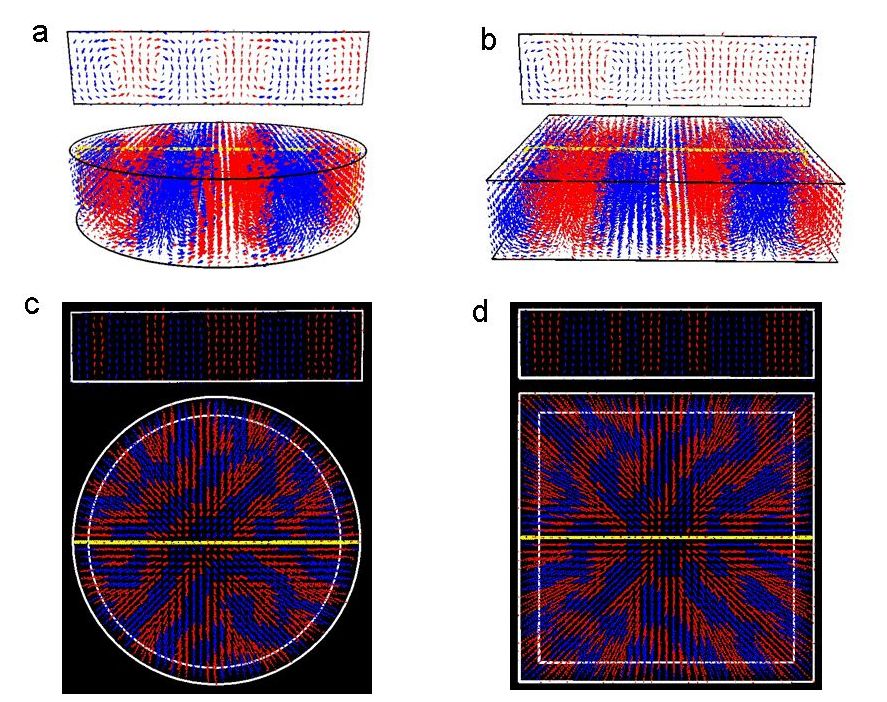}
\caption{Low-temperature three-dimensional polarization patterns in
free-standing (001) PZT (a,b) and BTO (c,d) nanoparticles under compressive
strain of -2$\%$. (a,c)-- 39$\times$10 nanodisks, (b,d)-- 39$\times$39$%
\times $10 particles with square footprints. The small rectangular inserts
show the cross sections of a central vertical plane indicated by yellow
lines. Red and blue arrows show local dipoles with positive and negative $z$%
-component, respectively. }
\label{fig:fig3}
\end{figure}

In the PZT nanoparticles, the 180$^{0}$ multi-domain phase is nothing but
the well-known 180$^{0}$ stripe domain structure found previously in
infinite PZT thin films \cite{wu1,bo1}. Here, like in the thin films, the
\textquotedblleft up" and \textquotedblleft down" domain run along [100]
 and alternate along [010] direction
(Fig.~3a,b). The domain width, however, is not homogeneous across the
particles: it is larger in the interior region and becomes 1-2 unit cell
narrower near the edge of the particles. Besides, the stripes become wider
in bigger particles. In a circular 6$\times $19 particle, for example, the
width of the central stripe is 4 unit cells (16 \AA ), while it is 6 unit
cells (24\AA ) in 10$\times $39 and 10$\times $39$\times $39 particles. In
both cases the stripes are narrower than that found in the PZT thin film
modeled with a 40$\times $40$\times $5 supercell (8 unit cells), despite the
fact that our particles are thicker.

\begin{figure}[tbp]
\centering
\includegraphics[scale=0.80]{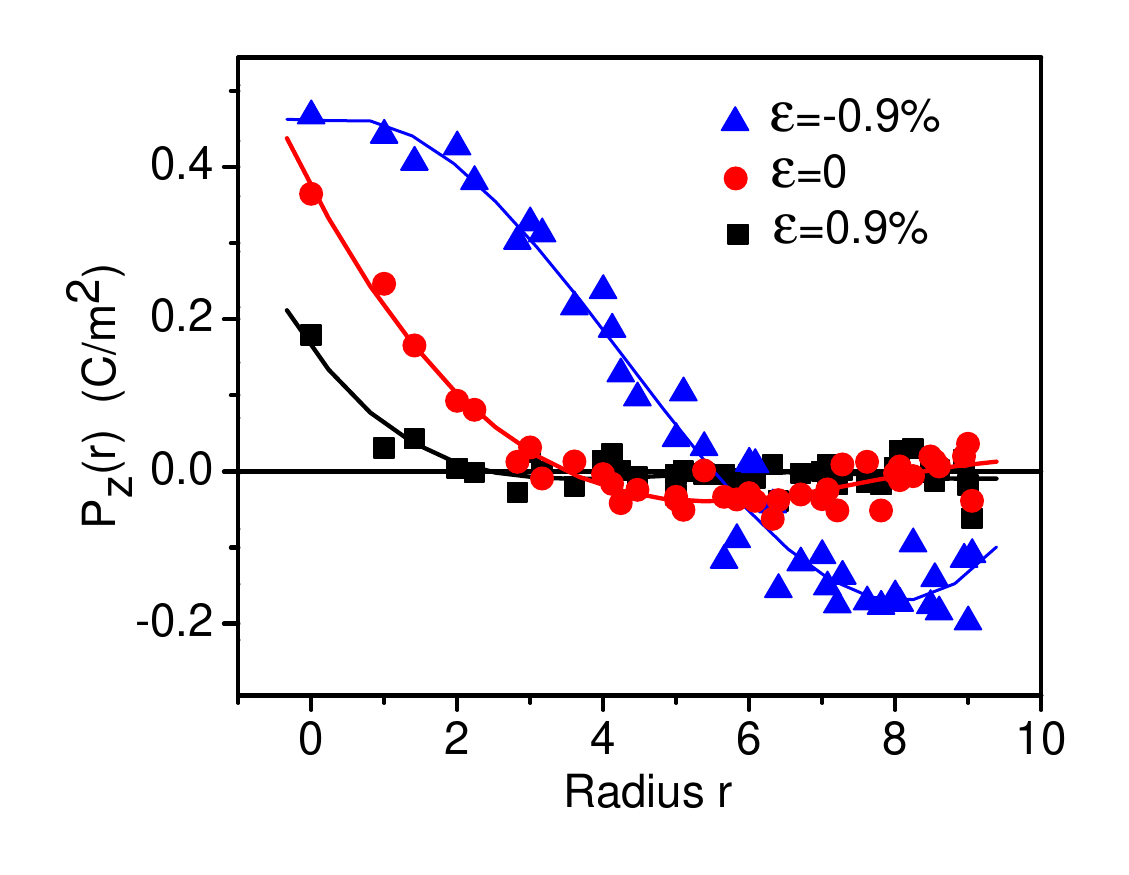}
\caption{Vertical polarization component in a 19$\times$6 PZT disk at
different misfit strains. While the values are radially averaged, they
correspond to an instant moment of time and represent the fluctuating dipole
structure. The lines represent 4-th order polinomial fits to the instant
polarizations.}
\label{fig:fig4}
\end{figure}

BTO nanoparticles, in comparison with PZT counterparts, adopt not straight,
but curved zig-zag-type stripes, running predominantly in [110] and [1-10]
directions (Fig.~3c,d). Though rarely, they intersect each other forming a
180$^{0}$ tweed domain texture, which is drastically different from that
found in BTO thin films \cite{tinte, bo2}. In the case of thin films, the
stripes run only along one direction (either [110] or [1-10]) and the system
loses its four-fold symmetry axis perpendicular to the film. In the
nanoparticles, however, the axis of the 4-th order is (on average) preserved
by allowing the forming stripes to propagate in both possible directions.
The stripes width changes from place to place, being less than 4 lattice
constants on average. This value can be compared with 4.3 lattice periods
found in BTO thin films \cite{bo2}.

As the present analysis shows, the transformation \textquotedblleft
vortex-to-180$^{0}$ domains" does not occur directly, but rather via one or
even two intermediate structures. The simplest case is presented by a 19$%
\times $6 PZT particle having only a single intermediate structure. This
structure develops from the vortex core that turns out to be longitudinally
polarized already in a free-standing state. While the core region (with the
radius 3-4 lattice spacings) embraces the polarization pointing
\textquotedblleft up", beyond this radius the polarization tilts weakly out
of plane oppositely to the core region, so that sum of all the dipoles add
up to zero (Fig.~4). Under compressive strains, the vortex core radius
expands and becomes comparable to the lateral size of the particles. At this
stage, the $x,y-$components of the local dipoles still form a vortex, while
the out-of-plain component, $z$, breaks the system into coaxial oppositely
polarized cylindrically symmetric domains--skyrmion-like structure \cite%
{roler}. As the compressive strain further increases, this structure
abruptly  transforms into a 180$^{0}$ stripe domain phase with zero moment $G$
(Fig.~2a).

\begin{figure}[tbp]
\centering
\includegraphics[height=35mm]{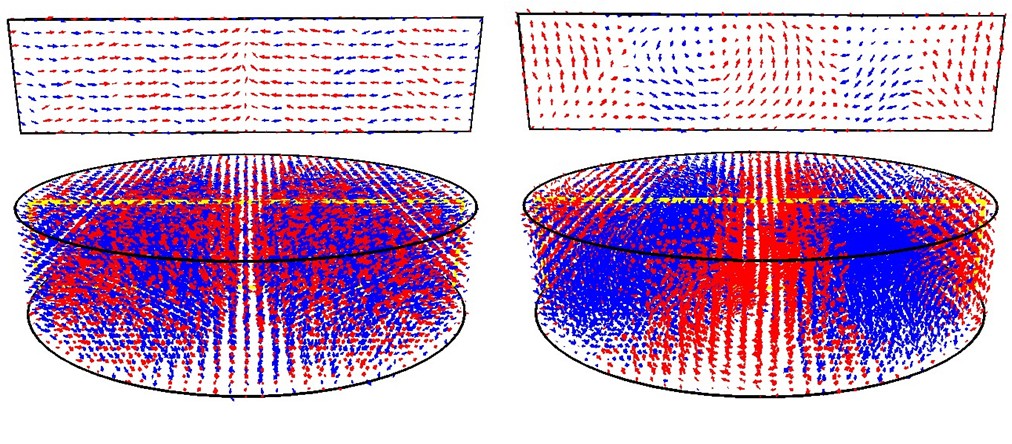}
\caption{Distribution of polarization in a 39$\times$10 PZT nanodisk for two
close strains of 0.6 $\%$ (left) and 0.5 $\%$ (right). The latter, which is
more compressive relative to the free standing nanoparticle, leads to a
state where 180$^0$ domains coexist with the curling polarization. The small
rectangular inserts show the cross sections of a central vertical plane
indicated by yellow lines. Red and blue arrows show local dipoles with
positive and negative $z$-component, respectively.}
\label{fig:fig5}
\end{figure}

In all other particles (both PZT and BTO), the structural transformation
into 180$^{0}$ domain phase occur in a two-step fashion. At first and
relatively short stage, again, two symmetry conforming 180$^{0}$ coaxial
domains develops at the expense of the toroidal moment (skyrmion structure).
As the in-plane compression increases, this skyrmion structure transforms
into 180$^{0}$ stripe (tweed) domains, as demostrated in Fig. 5 for a 39$%
\times $10 PZT nanoparticle. Remarkably, this happens on the background of
nonzero $G_{z}$. And finally, the toroidal moment disappears at a critical
compressive strain, and the system enters into a pure stripe or tweed domain
phase.

If the compressive strains are released, the system in a 180$^{0}$ domain
state relaxes back to a vortex dipole configuration. Typical evolution of
the toroidal moment and the average value of the $z$-component of the local
mode $|u_{z}|=N^{-1}\sum_{i}|u_{i,z}|$ with the number of the Monte-Carlo
sweeps is shown in Fig.~6. It is seen that the $G_{z}$ component first
linearly and then non-linearly increases, until saturation is reached, the other
two components being practically null. The $|u_{z}|$, which is simply
proportional to the average magnitude of the out-of-plane polarization $%
|P_{z}|$, decreases in a similar way, until it becomes saturated at some
non-zero value. The last fact is mainly connected with the oppositely
polarized core (\textquotedblleft up") and periphery region (\textquotedblleft down").

\begin{figure}[tbp]
\centering
\includegraphics[scale=0.75]{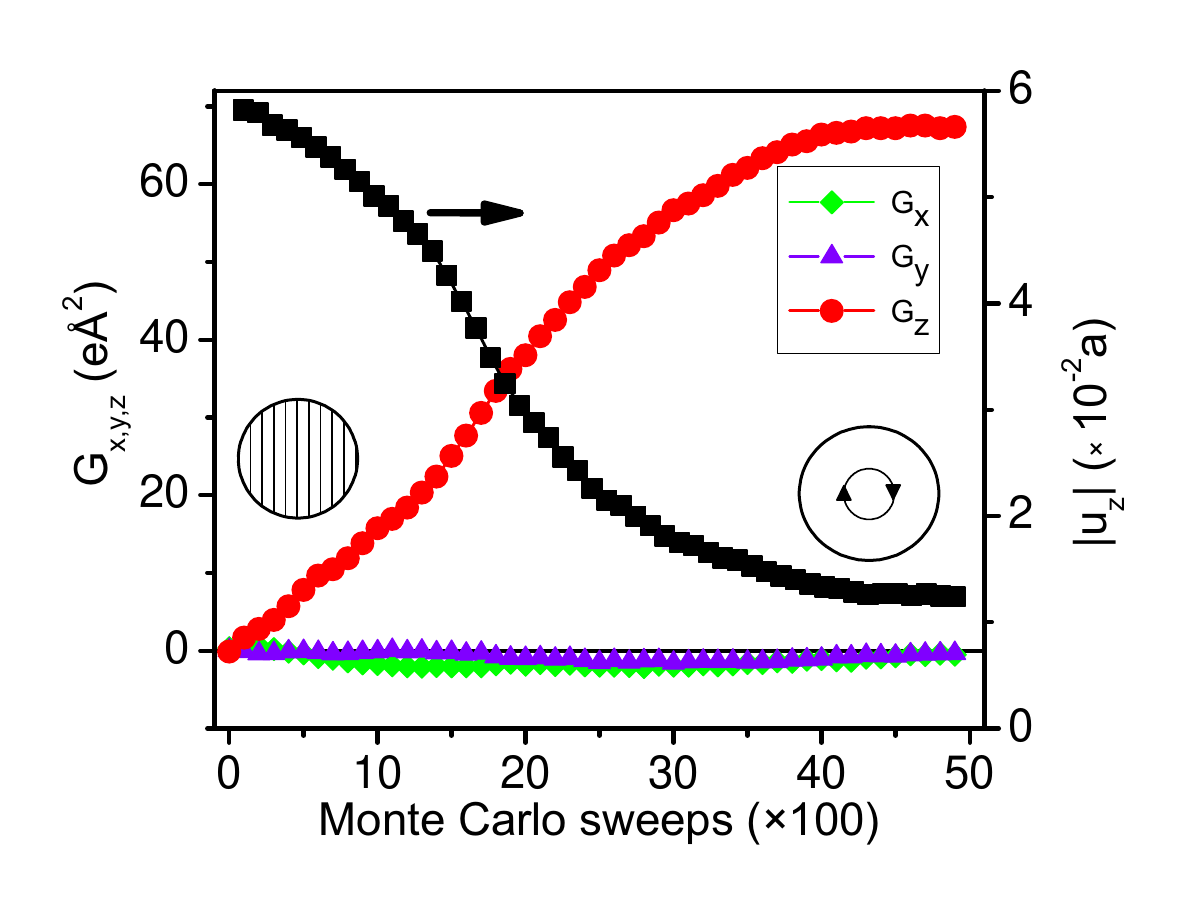}
\caption{The evolution of the toroidal moment $\mathbf{G}$ and average
magnitude of the z-component of the local mode in a 39$\times$10 PZT disk
after "switching" off compressive strains.}
\label{fig:fig6}
\end{figure}

\begin{figure*}[tbp]
\centering
\includegraphics[height=90mm]{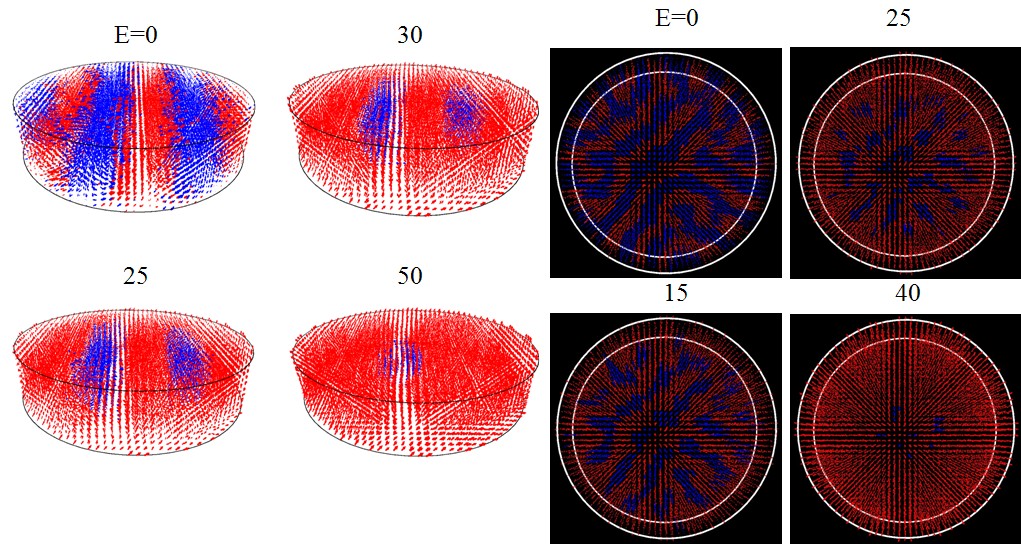}
\caption{Three-dimensional low-temperature polarization patterns in (001) 39$%
\times$10 PZT (white background) and BTO (black background) nanoparticles
under compressive strain of -2 $\%$, for different out-of-plane electric
fields (in 10$^6$ V/cm.}
\label{fig:fig7}
\end{figure*}

Consider now how the stripe domain and tweed structures evolve under the
external electric field. The typical behavior of PZT systems is presented by
a cylindrical particle 10$\times $39, whose evolution under different fields
applying along $z$ is shown in Fig.~7. On can see that the dipoles with the
direction opposite to the applied field start flipping first near the
particle edge. When the field is further increased, the \textquotedblleft
switching front" propagates to the center, shrinking the length of the
stripes and (to a lesser extent) their width, still being opposite to the
field. Ultimately, one reaches a state in which only one reduced stripe
remains, near the particle center; this stripe has an elliptical cross
section and can be called a \textquotedblleft bubble" (Fig.~7, $E=50\times
10^{6}$ V/cm). BTO nanoparticles behave qualitatively similar to the PZT
counterparts only in the initial stage of evolution: \ there the dipoles
also start flipping in the field direction in the rim region causing the
longitudinal shrinking of the stripes. Now, however, the stripe shrinking is
accompanied by their breaking into pieces (substripes) which, in turn,
transform into nanobubbles. As a result, several relatively thin bubbles
form near the center before the systems enters a monodomain state (Fig.~7, $%
40\times 10^{6}$ V/cm). Note that both in PZT and BTO, the out-of-plain
polarization linearly increases with the electric field, similar to the case
of ultra-thin films \cite{bo1, bo2}.

All the considered above intermediate phases (including skyrmion-like) are
characterized by $z$-domains coexisting with the in-plane curling
polarization; they can be called the  "domain-patterned vortex" structures.
Interestingly, such structures have never been reported for ferroelectric
small systems, though has been found in flat magnetic particles with low
perpendicular anisotropy \cite{maziewski}. Analogously, the 180$^{0}$ tweed
domain texture has never been discovered in ferroelectric nanostructures,
though resembles pretty much the labyrinthine phase observed in ultrathin
magnetic films \cite{portmann}. One can think that this structure is
metastable and eventually will evolve into an ordinary stripe phase. In our
simulations, however, we did not observe any signs of such transitions even
after a huge running time, more than 500 000 Monte Carlo sweeps.

Our predictions can be helpful for understanding some puzzling and
controversial experimental results \cite{roelofs,okaniwa,seol,shimizu}.
While the authors \cite{roelofs,okaniwa} did not observe any piezoresponse
of the PbTiO$_{3}$ islands with the diameters smaller than 20 nm, Ref. \cite%
{seol} reported observation of ferroelectricity in zirconate titanate
nanoparticles with the lateral size as small as 9 nm. On the other hand,
according to measurements \cite{shimizu}, the relationship between the size
of nanoislands and occurrence of ferroelectricity is rather random, raising
the question why some particles are ferroelectric, but the other (even with
very similar sizes) are not. Based on our results, we can explain these
uncertainties in the experimental results in the following way. As compared
with continuous thin films, the existence of free side walls in nanoislands
leads to new and additional mechanisms of the strain relaxations \cite%
{ovid'ko}, including formation of partial or full misfit dislocations at
nodes of the lateral surfaces and the flat plane of the substrate. This
implies that with the shrinking in lateral size, the maintaining of coherent
mismatch lattice strains in a nanoisland becomes more and more difficult,
and it is very likely that under the same experimental conditions the
smaller particles will experience less external strains. Moreover, since the
magnitude of strains is not well-controlled, even the particles with
comparable sizes can be differently strained. But in this case, according to
our calculations, they can show totally different piezoresponses. Indeed, as
the compressive strain is relaxed, an island can transform into an
intermediate or even in a vortex state, which is unable to produce any
(vertical or lateral) PFM signal at all. The particle in a vortex state
should be viewed as a paraelectric one showing only linear and
hysteresis-free piezoresponse as a function of dc bias voltage applied to
the AFM tip. Similar arguments can be used in interpretation the results of
Ref. \cite{chu} reporting the absence of the local piezoresponse loop in a
PZT nanoisland with a height of 10 nm \cite{chu}. Although the authors
attributed this effect to the inhomogeneous elastic fields associated with
misfit dislocations \cite{chu}, it also can be explained by the formation of
a curling polarization.

In summary, we investigate the strain effects on (001)-epitaxial PZT and
BTO nanoislands using first-principle-derived effective Hamiltonian
approach. Our study leads to the following conclusions: (i) regardless of
size, shape, and material, all the investigated particles transform into a
vortex state with in-plane curling polarization; the latter resembles
classical vortex pattern in smaller/round particles, while in
bigger/rectangular particles it can also be described as a domain structure
with four closure domains and four 90$^{0}$ domain walls, (ii) under strong
enough compressive strains the vortex state is no longer stable and yields
to a 180$^{0}$ stripe domain phase in PZT and to a more symmetric, 180$^{0}$
tweed domain texture, in BTO nanoparticles, and (iii) these transitions
proceed not directly, but via one or two \textit{unusual} intermediate
phases where domain structure coexist with circular vortex ordering. We
further discovered that the electric-field induced evolution (switching) of
multi-domain structures is essentially different for the two materials (PZT
and BTO).

\end{document}